# Chaotic Encryption Scheme Using 1-D Chaotic Map

**Mina Mishra, Vijay H. Mankar**
Ph.D. Scholar (*Electronics & Telecommunication*), *Nagpur University*, *Maharastra*, *India*
*Senior Faculty*, *Department of Electronics Engineering*, *Government Polytechnic Nagpur*, *Maharastra*, *India*
*E-mail*: {*minamishraetc*, *vhmankar*}@gmail.com

## Abstract

This paper proposes three different chaotic encryption methods using 1-D chaotic map known as Logistic map named as Logistic, NLFSR and Modified NLFSR according to the name of chaotic map and non-linear function involved in the scheme. The designed schemes have been crypt analyzed for five different methods for testing its strength. Cryptanalysis has been performed for various texts using various keys selected from domain of key space. Logistic and NLFSR methods are found to resist known plaintext attack for available first two characters of plaintext. Plaintext sensitivity of both methods is within small range along with medium key sensitivity. Identifiability for keys of first two of the scheme has not been derived concluding that methods may prove to be weak against brute-force attack. In the last modified scheme avalanche effect found to be improved compared to the previous ones and method is found to resist brute-force attack as it derives the conclusion for identifiability.

## 1. Introduction

Chaotic cryptography [1] deals with hiding and recovering of secret messages using algorithm which consists of encryption rule that uses chaotic functions (analog or digital). Logistic map is one-dimensional map that consists of a parameter which is acting as secret key in the designed encryption schemes. The proposed methods use non-linear functions like, sinusoidal, non-linear shift register and logistic map to built confusion and diffusion.

The sine function describes a smooth repetitive oscillation. It's most basic form as a function of time (*t*) is:

$$y(t) = A \sin(wt + \phi)$$

where, *A*, the amplitude is the peak deviation of the function from its center position. *ω*, the *angular frequency*, specifies how many oscillations occur in a unit time interval, in radians per second. *φ*, the *phase*, specifies where in its cycle the oscillation begins at *t* = 0.

A NLFSR (Non-Linear Feedback Shift-register) is a common component in modern stream ciphers, especially in RFID and smartcard applications. NLFSRs are known to be more resistant to cryptanalytic attacks than Linear Feedback Shift Registers (LFSR's), although construction of large NLFSRs with guaranteed long periods remains an open problem. A NLFSR, is a shift register whose current state is a non-linear function of its previous state. The NLFSR used here is shown in **Figure 1** output.

The logistic map is a polynomial mapping of degree 2, it takes a point, in a plane and maps it to a new point using following expressions:

$$x(k+1) = r\, x(k)\,(1 - x(k));$$

where, map depends on the parameter *r*. From *r* = 3.57 to *r* = 4, the map exhibits chaotic behavior which is shown in **Figure 2**.

Cryptanalysis is the study of attacks against cryptographic schemes to disclose its possible weakness. During cryptanalyzing a ciphering algorithm, the general assumption made is that the cryptanalyst knows exactly the design and working of the cryptosystem under study, *i.e.*, he/she knows everything about the cryptosystem except the secret key. It is possible to differentiate between different levels of attacks on cryptosystems. They are briefly explained as follows:

a) Cipher text-only attack: The attacker possesses a string of cipher text.

b) Known plain text: The attacker possesses a string of plain text, p, and the corresponding cipher text, c.

c) Chosen plain text: The attacker has obtained temporary access to the encryption machinery. Hence he/she can choose a plain text string, p, and construct the corre

 

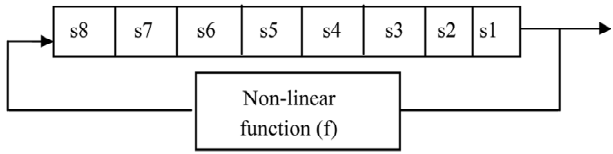

**Figure 1. NLFSR using 8-bit shift registers.**

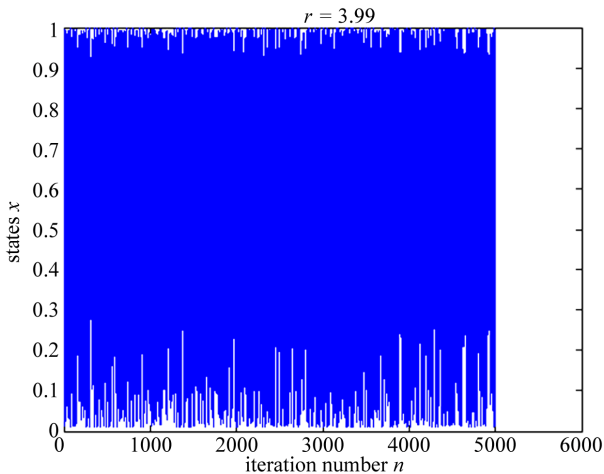

**Figure 2. Plot of Logistic map for $r = 3.99$, $x(0) = 0.99$, $n = 5000$.**

sponding cipher text string, c.

d) Chosen cipher text: The attacker has obtained temporary access to the decryption machinery. Hence he/ she can choose a cipher text string, c, and construct the corresponding plain text string, p.

e) Brute Force Attack: A brute force attack is the method of breaking a cipher by trying every possible key. The brute force attack is the most expensive one, owing to the exhaustive search.

In addition to the five general attacks described above, there are some other specialized attacks, like, differential and linear attacks.

Differential cryptanalysis is a kind of chosen-plaintext attack aimed at finding the secret key in a cipher. It analyzes the effect of particular differences in chosen plaintext pairs on the differences of the resultant cipher text pairs. These differences can be used to assign probabilities to the possible keys and to locate the most probable key.

Linear cryptanalysis is a type of known-plaintext attack, whose purpose is to construct a linear approximate expression of the cipher under study. It is a method of finding a linear approximation expression or linear path between plaintext and cipher text bits and then extends it to the entire algorithm and finally reaches a linear approximate expression without intermediate value.

## 2. Methodology

The complete methodology involved during designing is clearly cited with the help of block diagram as shown in **Figure 3**.

The plain text is encrypted by an encryption rule which uses non-linear function and the state generated by the chaotic system in the transmitter [2]. The scrambled output is inputted further to the chaotic system such that the chaotic dynamics is changed continuously in a very complex way. Then another state variable of the chaotic system [3] in the transmitter is transmitted through the channel.

Recovery of the plaintext is done by decrypting the input (ciphertext) using reverse process of encryption, as used in the transmitter.

In the modified NLFSR method traditional encryption method is used along with above method. For Logistic-sinusoidal function and logistic map and for NLFSR-Non-linear shift register and logistic map has been used respectively.

## 3. Analysis and Results

The analysis accomplished on the designed ciphers has been done for key space [4], avalance effect, known-plaintext attack, Identifiability [5] [6] and results are cited in tabular form.
  a) Logistic: Key Space range is from 3.57 to 4.0;
  b) NLFSR: Key Space range is from 3.57 to 4.0;
  c) Modified NLFSR: Key space is from 3.57 to 4.0.

## 4. Conclusions and Future Scope

**Tables 1** and **2** show that Logistic and NLFSR methods are found to resist known plaintext attack for available first two characters of plaintext but modified NLFSR proves its resistance against the attack for available some numbers of characters of plaintext as given in the column of **Table 3**. It also proves its resistance against brute-force attack by deriving conclusion for identifiability of selected key. All the three methods are similar to one-time pad type instead more secure.

More secure ciphers can be designed and crypt analyzed [7,8] using different non-linear functions and 2-D

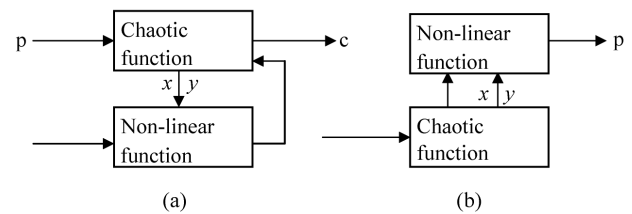

**Figure 3. Methodology used in design of chaotic cryptosystem. (a): Transmitter end (encryptor); (b): Receiver end (decryptor). (p: plaintext; c: ciphertext; $x$: state of chaotic function; $y$: Intermediate encrypted plaintext).**

**Table 1. Analysis Table for Logistic method.**

| Sl. no. | Plaintext | Key value | Ciphertext | Plaintext sensitivity (in %) | Key sensitivity (in %) | Domain for key with increment = 0.0001 | Identifiability of key for iteration value = 1 or 2 | Robustness against known plaintext attack for p=[$p_1$ $p_2$] | Whether key can act as secret key against Brute Force attack? |
|---|---|---|---|---|---|---|---|---|---|
| 1. | Hello! how are you? | 3.65 | BHZOR0Nhw`X Vb?WoM | 3.9474 | 46.7105 | (3.57, 3.78) | NI | R | NO |
| 2. | Ram scored 98 marks in Maths. | 3.71 | jPw+4(f}mmWek f8bq9m C|*i: | 2.0833 | 43.7500 | (3.67, 3.87) | NI | R | NO |
| 3. | Thank you,sir. | 3.87 | >3ieVyl1[w5f\ | 4.1667 | 50 | (3.77, 3.97) | NI | R | NO |
| 4. | The match was very exciting. | 3.88 | J?/el?jZl67 H3;L LE~2Je-yeOzY | 4.7414 | 46.1207 | (3.78, 3.98) | NI | R | NO |
| 5. | I will be leaving at 9 p.m. | 3.89 | ZDWuQ/9H<r4c* \{>l}s?D4UgX | 1.8519 | 42.5926 | (3.79, 3.99) | NI | R | NO |

NI – Non-Identifiable ; I – Identifiable ; R – Robust; p [$p_1$ $p_2$…$p_n$] – First 'n' characters of available plaintext string.

**Table 2. Analysis table for NLFSR method.**

| Sl.no. | Plaintext | Key value | Ciphertext | Plaintext sensitivity (in %) | Key sensitivity (in %) | Domain for key With increment = 0.0001 | Identifiability of key for iteration value =1 or 2 | Robustness against known plaintext attack for p = [$p_1$ $p_2$] | Whether key can act as secret key against Brute Force attack? |
|---|---|---|---|---|---|---|---|---|---|
| 1. | I am going to market. | 3.7328 | ¸èøxè0ø¸OØ¨0v | 0.5682 | 20.4545 | (3.57,3.77) | NI | R | NO |
| 2. | Hello!how are you? | 3.7694 | ¨78÷øðP¨ø°ý | 0.6579 | 25 | (3.57,3.77) | NI | R | NO |
| 3. | Ram scored 98 marks in Maths. | 3.8551 | K¸ÐÈ÷P¨'¸PØÏw ´0Ïv | 0.4167 | 18.3333 | (3.76,3.96) | NI | R | NO |
| 4. | Thank you,sir. | 3.8641 | ¸w× ø¯ 6ÐPu | 0.8333 | 30.8333 | (3.78,3.98) | NI | R | NO |
| 5. | The match was very exciting. | 3.9065 | +§·0ÇïÏo§P§- È/wèu | 0.4310 | 28.0172 | (3.785,3.985) | NI | R | NO |

**Table 3. Analysis table for modified NLFSR method.**

| Sl. no. | Plaintext | Key value | Ciphertext | Plaintext sensitivity (in %) | Key sensitivity (in %) | Domain for key With increment = 0.0001 | Identifiability of key for iteration value = 1 or 2 | Robustness against known plaintext attack. | Whether key can act as secret key against Brute Force attack? |
|---|---|---|---|---|---|---|---|---|---|
| 1. | What is your name? | 3.6424 | -§Án_ªÏ¾X<Ó¬ sÊmÄ | 9.2105 | 16.4474 | (3.57,3.77) | I | R for p=[$p_1$ $p_2$ …$p_5$] | YES |
| 2. | I am going to market. | 3.7328 | ‖ vfª(ò`(¬ô6ýR2 & | 10.7955 | 17.0455 | (3.57,3.77) | I | R for p=[$p_1$ $p_2$…$p_{19}$] | YES |
| 3. | Sita is singing very well. | 3.8544 | BÛ¼~eÂ=ßRf1 àdØùFì;d*p | 6.4815 | 12.9630 | (3.66,3.86) | I | R for p=[$p_1$ $p_2$ …$p_{19}$] | YES |
| 4. | Ram scored 98 marks in Maths. | 3.8551 | Lª ½jdSÜ* ¢æ %<Ãc&;$ÍèÅ<- ±& | 5.8333 | 14.5833 | (3.76,3.96) | I | R for p=[$p_1$ $p_2$ …$p_{19}$] | YES |
| 5. | Jaycee publication. | 3.8529 | =txý®&Ó`Fà- â$# | 14.3750 | 23.1250 | (3.77,3.97) | I | R for p=[$p_1$ $p_2$ … $p_{15}$] | YES |

chaotic maps [9] using different encryption rule. Improvement in the property of avalanche effect of designed ciphers is required in future.